\documentclass[12pt]{article}
\usepackage{a4wide}
\usepackage{epsfig}
\usepackage{amsmath}

\newlength{\absize}
\catcode`@=11
\def\citer{\@ifnextchar [{\@tempswatrue\@citexr}{\@tempswafalse\@citexr[]}}

\def\@citexr[#1]#2{\if@filesw\immediate
  \write\@auxout{\string\citation{#2}}\fi
  \def\@citea{}\@cite{\@for\@citeb:=#2\do
    {\@citea\def\@citea{--\penalty\@m}\@ifundefined
       {b@\@citeb}{{\bf ?}\@warning
       {Citation `\@citeb' on page \thepage \space undefined}}%
\hbox{\csname b@\@citeb\endcsname}}}{#1}} \catcode`@=12

\begin{document}

\begin{titlepage}

\begin{flushright}

\end{flushright}

\vspace{1truecm}

\begin{center}

{\Large \bf Representation of Noncommutative Phase Space }

\vspace{2cm} Kang Li$^a$\footnote{kangli@hztc.edu.cn},
 Jianhua Wang$^b$  and  Chiyi Chen$^c$

\vspace{1cm}

$^a${\it\small Department of Physics, Hangzhou Teacher's College,
Hangzhou, 310036, P.R. China\\
$^b$  Shaanxi College of Science and Engineer, Hanzhong,
723000,P.R. China\\
$c$ Shanghai Astronomical Observatory, Chinese Academy of
Sciences, Shanghai 200030, PR China

\smallskip }

\vspace{1cm}

\begin{abstract}
\noindent The representations of the algebra of coordinates and
momenta of noncommutative phase space are given. We study, as an
example, the harmonic oscillator in noncommutative space of any
dimension. Finally the  map of Schr$\ddot{o}$dinger equation from
noncommutative space to commutative space is obtained.

\bigskip

\end{abstract}

\end{center}
\vskip 0.5truecm

\end{titlepage}
\newpage
\baselineskip 6 mm


Recently, there has been much interest in the study of physics in
noncommutative space (NCS)\citer{SW,Sch}, not only because the NCS
is necessary when one studies the low energy effective theory of
D-brane with B field background, but also because in the very tiny
string scale or at very high energy situation, the effects of non
commutativity of space may appear. In literature the
noncommutative quantum mechanics and noncommutative quantum field
theory have been studied extensively, and the main approach  is
based on the Weyl-Moyal correspondence which amounts to replacing
the usual product by a star product in non-commutative space.

In this paper, rather than studying the star product, we analyze
the noncommutative effects in the usual  quantum mechanics phase
space. The usual method to do this is to use the Seiberg-Witten
(SW) type map\cite{SW}, that is to map the Gauge field or energy
momentum tensor to their counter-part in commutative space. This
method has been developed a lot in the past few years\cite{BLY}.
Our aim in this paper is to give a convenient method, albeit
equivalent to the SW map, to study the noncommutative effects
using commutative space coordinates. First we will give a
representation of the noncommutative coordinates and
noncommutative momenta in terms of the commutative coordinates and
momenta of usual quantum mechanics. Then, as an example, we study
the noncommutative harmonic oscillator in any dimension. At last,
we give the representation of the Hamiltonian operator of
noncommutative space and, as a consequence, the
Schr$\ddot{o}$dinger equation containing the noncommutative
effects is obtained.

In the usual commutative space (say n dimensional space), the
coordinates and momenta in quantum mechanics have the following
commutation relations:

\begin{eqnarray}
\label{Eq:cmr} \left\{
 \begin{array}{l}
~[x_{i},x_{j}]=0,\\~ [p_{i},p_{j}]=0,\\~ [x_{i},p_{j}]=i
\hbar\delta_{ij}.
\end{array}
\right.
\end{eqnarray}
At very tiny scales, say string scale, the space may not commute
anymore. Let us denote the operators of coordinates and momenta in
noncommutative space as $\hat{x}$ and $\hat{p}$ respectively, then
the $\hat{x}_i$ and $\hat{p}_i$ will have the following algebra
relations, if both space-space and momentum-momentum
non-commutativities are considered.
\begin{equation}
\label{Eq:nmr} \left\{
 \begin{array}{l}
~[\hat{x}_{i},\hat{x}_{j}]=i\hbar\Theta_{ij}, \\~
[\hat{p}_{i},\hat{p}_{j}]=i\hbar\bar{\Theta}_{ij},\\~
[\hat{x}_{i},\hat{p}_{j}]=i \hbar\delta_{ij}.
\end{array}
\right.
\end{equation}
where $\{\Theta_{ij}\}$ and $\{\bar{\Theta}_{ij}\}$  are totally
antisymmetric matrices with very small elements representing the
noncommutative property of the space and momentum in
noncommutative phase space. Using the equation (\ref{Eq:cmr}) and
(\ref{Eq:nmr}), we can study the representations of the
noncommutative $\hat{x}$ and $\hat{p}$ in term of $x$ and $p$.
Once these representations are obtained, then the noncommutative
problems can be changed into problems in the usual commutative
space which we are familiar with.  In order to do so, let us give
an ansatz as follows (the summation between repeated indices is
implied in this paper ).
\begin{equation}
\label{Eq:rep.1}\left\{
 \begin{array}{ll}
 ~\hat{x}_{i}&=a_{ij}x_{j}+b_{ij}p_{j},\\~
 \hat{p}_{i}&=c_{ij}x_{j}+d_{ij}p_{j}.\\~
 ~& i,j= 1,2 ... n
\end{array}
\right.
\end{equation}
In matrix form, we have
\begin{equation}
\label{Eq:rep.2}\left(
 \begin{array}{l}
 \hat{x}  \\
 \hat{p}
\end{array}
\right)=\left(
 \begin{array}{ll}
 A & B \\
 C & D
\end{array}
\right)\left(
 \begin{array}{l}
 x \\
 p
\end{array}
\right)
\end{equation}
where $A=\{a_{ij}\}, B=\{b_{ij}\} ,C=\{c_{ij}\}$ and
$D=\{d_{ij}\}$ are $n\times n$ matrices.

Now let us calculate the exact form of the representation
matrices. For this, we insert equation (\ref{Eq:rep.1}) into
equation (\ref{Eq:nmr}), and using equation (\ref{Eq:cmr}) we
obtain
\begin{equation}
\label{Eq:Rel.}
\begin{array}{l}
AB^{T}-BA^{T}=\Theta \\
CD^{T}-DC^{T}=\bar{\Theta}\\
AD^{T}-BC^{T}= \ \mathbf{1}
\end{array}
\end{equation}
where $\mathbf{1}$ is the identity matrix and the upper index $T$
of a matrix means its transpose. In the above relations, the
matrices $A$ and $D$ could be chosen as proportional to identity
(scaling factors), so we denote them by $\alpha$ and $\beta$
respectively\footnote{Symmetry considerations tell us that these
two constant matrices should be same. This will be shown to be the
case in our generic discussion of noncommutative harmonic
oscillator in n-dimensions in the later part of this paper}. Then
the relations above change into,
\begin{equation}
\label{Eq:Rel.2}
\begin{array}{l}
\alpha ( B^{T}-B )=\Theta \\
\beta (C-C^{T})=\bar{\Theta}\\
BC^{T}= (\alpha\beta-1)\mathbf{1}
\end{array}
\end{equation}

From first two equations of (\ref{Eq:Rel.2}), we can see that if
$B$ and $C$ are antisymmetric, then they will have explicit
solutions, in fact if we assume $B$ and $C^T$ are commuting, then
the third equation of the (\ref{Eq:Rel.2}) is satisfied when $ B$
and $C$ are either symmetric or antisymmetric. The case of $B$ and
$C$ being both symmetric leads to the $\Theta = \bar{\Theta} =0$
which means our space is commutative. Thus, we choose $B$ and $C$
as antisymmetric matrices. Then we obtain
\begin{equation}
\label{Eq:Res.1}
\begin{array}{l}
B =-\frac{1}{2\alpha}\Theta ,\\
~\\
 C=~\frac{1}{2\beta}\bar{\Theta}.
\end{array}
\end{equation}
Here the scaling constants $\alpha$ and $\beta$ should not be
equal to zero.  The last equation of (\ref{Eq:Rel.2}) gives the
relationship between $\bar{\Theta}$ and $\Theta$
\begin{equation}
\label{TB}
 \Theta\bar{\Theta}=4\alpha\beta (\alpha\beta-1)\cdot\mathbf{1}.
\end{equation}

One can then put the general representation matrix of
$(\hat{x},\hat{p})$ in the commutative space as:
\begin{equation}
\label{Eq:Rep.3} \left(
\begin{array}{ccc}
 \alpha\cdot\mathbf{1}&~&-\frac{1}{2\alpha}\Theta ,\\
 \frac{1}{2\beta}\bar{\Theta}&~&\beta\cdot\mathbf{1} \end{array}\right)_{2n\times
 2n}.
\end{equation}
And equations (\ref{Eq:rep.1}) become,
\begin{equation}
\label{Eq:Rep.4}\left\{
 \begin{array}{ll}
 \hat{x}_{i}&= \alpha x_{i}-\frac{1}{2\alpha}\Theta_{ij}p_{j},\\
 ~&~\\
 \hat{p}_{i}&=\beta p_{i}+\frac{1}{2\beta}\bar{\Theta}_{ij}x_{j} .\\
 ~&~\\
 ~& i,j= 1,2 ... n
\end{array}
\right.
\end{equation}
When $\alpha=\beta=1$, it will correspond to $\bar{\Theta}=0$ (
refer to the next part of this paper), which is the case that is
extensively studied in the literature where the space coordinates
are non-commuting while momentum space is commuting.

The angular momentum of noncommutative space is given by,
\begin{equation}
\hat{L}_{i}=\epsilon_{ijk}\hat{x}_{j}\hat{p}_{k}.
\end{equation}
By using the representation equations (\ref{Eq:Rep.4}), the
$\hat{L}$ in usual commutative space can be written as,
\begin{equation}
\label{Eq:am1}
 \begin{array}{ll}
 \hat{L}_{i}=& \epsilon_{ijk}( \alpha\beta
 x_{j}p_{k}+\frac{\alpha}{2\beta}x_{j}\bar{\Theta}_{km}x_{m}\\
 ~&
 -\frac{\alpha}{2\beta}\Theta_{jl}p_{l}p_{k}-\frac{1}{4\alpha\beta}\Theta_{jl}p_{l}\bar{\Theta}_{km}x_{m}).
\end{array}
\end{equation}
Or we can write it in vector form ,
\begin{equation}
\label{Eq:am2}
 \begin{array}{ll}
 \hat{\mathbf{L}}=& \alpha\beta \mathbf{x}\times \mathbf{p} -\frac{\alpha}{2\beta}(\Theta \mathbf{p} )\times\mathbf{p
 }\\
 ~&
 +\frac{\alpha}{2\beta}\mathbf{x}\times (\bar{\Theta}\mathbf{x})

-\frac{1}{4\alpha\beta} (\Theta \mathbf{p})\times
(\bar{\Theta}\mathbf{x}).
\end{array}
\end{equation}
And the commutation relations between $\hat{\mathbf{L}}$ and
$\hat{\mathbf{x}},\hat{\mathbf{p}}$ have the forms as follows,
\begin{equation}
\label{Eq:cr1}
 \begin{array}{ll}
 [\hat{L}_{i},\hat{x}_{j}]&=[\epsilon_{ikl}\hat{x}_{k}\hat{p}_{l},\hat{x}_{j}]\\
                         ~&=i\hbar\epsilon_{ijk}\hat{x}_{k}+i\hbar\epsilon_{ikl}\Theta_{kj}\hat{p}_{l},
\end{array}
\end{equation}
\begin{equation}
\label{Eq:cr2}
 \begin{array}{ll}
 [\hat{L}_{i},\hat{p}_{j}]&=[\epsilon_{ikl}\hat{x}_{k}\hat{p}_{l},\hat{p}_{j}]\\
                         ~&=i\hbar\epsilon_{ijk}\hat{p}_{k}-i\hbar\epsilon_{ikl}\bar{\Theta}_{kj}\hat{x}_{l},
\end{array}
\end{equation}

Up to now, we have given the representation of coordinates and
momenta of noncommutative space in terms of the coordinates and
momenta in the usual quantum mechanics commutative space. Using
these representations we can study the effects of
non-commutativity in the usual space we are familiar with. As an
illustrative example, let us study in detail the harmonic
oscillator in noncommutative space with arbitrary  dimensions $n$.
In $n$ dimensional noncommutative space, the Hamiltonian of the
harmonic oscillator has the form as,
\begin{equation}
\label{H1}
\hat{H}=\frac{1}{2\mu}\hat{p_{i}}\hat{p}_{i}+\frac{1}{2}\mu\omega^2\hat{x}_{i}\hat{x}_{i}
\end{equation}
where the $\mu$ and $\omega$ represent the mass and angular
frequency of the oscillator. Using the representation equations in
(\ref{Eq:Rep.4}), we obtain the expression of the $\hat{H}$ in the
commutative space as,
\begin{equation}
\label{Eq:H}
 \begin{array}{ll}
 \hat{H} &= \frac{\beta^2}{2\mu}p_i p_i+\frac{\alpha^2}{2}\mu\omega^2
 x_i x_i +(\frac{1}{4}\mu\omega^2
       \Theta_{ij}+\frac{1}{4\mu}\bar{\Theta}_{ij})(p_i x_j +x_j p_i)\\
       ~& +\frac{\mu\omega^2}{8\alpha^2}\Theta_{ij}\Theta_{il}p_j
       p_l
       +\frac{1}{8\mu\beta^2}\bar{\Theta}_{ij}\bar{\Theta}_{il}x_j
       x_l
\end{array}
\end{equation}
Again we can see that when $\alpha=\beta=1$, corresponding to
$\bar{\Theta}_{ij}=0$, then the above Hamiltonian corresponds to
the case where the space is non-commutative while the momenta are
commuting. If we set further more  $\Theta_{ij}=0$ then the result
corresponds to the usual $n$ dimension harmonic oscillator in
commutative space.

For convenience, the annihilation operators $a$ and creation
$a^\dag$ operators are introduced when one studies harmonic
oscillator problem. Similarly, we would like to introduce their
counterparts $\hat{a}$ and $\hat{a}^\dag$ in noncommutative space
and the relationship between them.  The deformed
annihilation-creation operators in $n$-dimensional noncommutative
space are defined by
\begin{equation}
\label{Eq:aa+}
\hat{a}_{i}=\sqrt{\frac{\mu\omega}{2}}(\hat{x}_{i}+\frac{i}{\mu\omega}\hat{p}_{i})~~~~
\hat{a}^{\dag}_{i}=\sqrt{\frac{\mu\omega}{2}}(\hat{x}_{i}-\frac{i}{\mu\omega}\hat{p}_{i})
\end{equation}
It is easy to check that,
\begin{equation}
\label{Eq:aa+2}
 \begin{array}{l}
~[\hat{a}_{i},\hat{a}^{\dag}_{j}]=\delta_{ij}+i\mu\omega\Theta_{ij}\\
~[\hat{a}_i,\hat{a}_j]=[\hat{a}^{\dag}_i,\hat{a}^{\dag}_j]=\frac{i}{2}\mu\omega
(\Theta_{ij}-\frac{1}{\mu^{2}\omega^{2}}\bar{\Theta}_{ij})
\end{array}
\end{equation}
In order to keep Bose-Einstein statistics in noncommutative case
we need $\hat{a}^\dag_i$ and $\hat{a}^\dag_j$ to be commuting,
whence the consistency condition:
\begin{equation}
\label{Eq:cc}
 \bar{\Theta}=\mu^2\omega^2\Theta
\end{equation}
Under this condition, the above commutators become
\begin{equation}
\label{Eq:aa+3}
 \begin{array}{l}
~[\hat{a}_{i},\hat{a}^{\dag}_{i}]= 1,\\
~[\hat{a}_i,\hat{a}_j]=[\hat{a}^{\dag}_i,\hat{a}^{\dag}_j]=0.
\end{array}
\end{equation}
which are just the same as the ones in commutative space. But
there is also an extra commutator
\begin{equation}
\label{Eq:aa+4}
[\hat{a}_{i},\hat{a}^{\dag}_{j}]=i\mu\omega\Theta_{ij}, {\rm
for}~~ i\neq j .
\end{equation}
We point out here that this abnormal extra commutator is the
reason of the existence of fractionally angular
momentum\cite{zhang}, and it is consistent with all principles and
Bose-Einstein statistics. Replacing (\ref{Eq:cc}) in
(\ref{Eq:Rep.4}) we have
\begin{equation}
\label{Eq:Rep.5}\left\{
 \begin{array}{ll}
 \hat{x}_{i}&= \alpha x_{i}-\frac{1}{2\alpha}\Theta_{ij}p_{j},\\
 ~&~\\
 \hat{p}_{i}&=\beta p_{i}+\frac{1}{2\beta}\mu^2\omega^2\Theta_{ij}x_{j}.\\
 ~&~\\
 ~& i,j= 1,2 ... n
\end{array}
\right.
\end{equation}

As we know in the commutative space,  the annihilation and
creation operators can be expressed as
\begin{equation}
\label{Eq:aa+c}
a_{i}=\sqrt{\frac{\mu\omega}{2}}(x_{i}+\frac{i}{\mu\omega}p_{i}),~~~~
a^{\dag}_{i}=\sqrt{\frac{\mu\omega}{2}}(x_{i}-\frac{i}{\mu\omega}p_{i}).
\end{equation}
So from a straight calculation we get the relation between
$\hat{a}_i , \hat{a}^\dag_i$ and $a_i , a^\dag_i$,
\begin{equation}
\label{Eq:aa+r}
\begin{array}{ll}
\hat{a}_{i}=&\frac{1}{2}(\alpha +\beta)a_i +\frac{i}{4}\mu\omega
          (\frac{1}{\beta}+\frac{1}{\alpha})\Theta_{ij}a_j\\
          ~&+\frac{1}{2}(\alpha -\beta)a^\dag_i +\frac{i}{4}\mu\omega
          (\frac{1}{\beta}-\frac{1}{\alpha})\Theta_{ij}a^\dag_j ,\\
          ~&~\\
\hat{a}^{\dag}_{i}=&\frac{1}{2}(\alpha
+\beta)a^\dag_i-\frac{i}{4}\mu\omega
          (\frac{1}{\beta}+\frac{1}{\alpha})\Theta_{ij}a^\dag_j\\
            ~&+\frac{1}{2}(\alpha -\beta)a_i
            -\frac{i}{4}\mu\omega
          (\frac{1}{\beta}-\frac{1}{\alpha})\Theta_{ij}a_j .\\
\end{array}
\end{equation}
Causality considerations tell us that if a  particle is
annihilated in noncommutative space, and in order to view this
phenomenon in commutative space, then it should correspond to some
sort of annihilation in the latter space. That is to say
$\hat{a}_i$ should only be some combination of $a_i$'s. For the
same reason, $\hat{a}^\dag_i$ should also be some combination of
$a^\dag_i$'s. So this forces us to set,
\begin{equation}
\label{alpha} \alpha = \beta =: \alpha .
\end{equation}
 And then the equations in  (\ref{Eq:aa+r}) become
\begin{equation}
\label{Eq:aa+r2}
\begin{array}{ll}
\hat{a}_{i}=& \alpha  a_i +\frac{i}{2\alpha}\mu\omega\Theta_{ij}a_j,\\
\hat{a}^{\dag}_{i}= & \alpha a^\dag_i-\frac{i}{2\alpha}\mu\omega
          \Theta_{ij}a^\dag_j .
\end{array}
\end{equation}
Now we can discuss the Hamiltonian and angular momentum in term of
annihilation and creation operators. It is easy to check that the
Hamiltonian of harmonic oscillator in equation (\ref{H1}) has the
form
\begin{equation}
\label{H2}
\begin{array}{ll}
\hat{H}&=\hbar\omega \hat{a}^\dag_i \hat{a}_i
+\frac{n}{2}\hbar\omega \\
~&=\hbar\omega [\alpha^2 a^\dag_i a_i + i\mu\omega
\Theta_{ij}a^\dag_i a_j +
\frac{1}{4\alpha^2}\mu^2\omega^2\Theta_{ij}\Theta_{il}a^\dag_j a_l
+\frac{n}{2}].
\end{array}
\end{equation}
We can see that the vacuum energy (the last term) exists also in
the noncommutative case, while the non-commutative effects appear
in the second and third terms related to the $\Theta$. The angular
momentum can be written as

\begin{equation}
\label{L}
\begin{array}{ll}
\hat{L}_i&=\epsilon_{ijk}\hat{x}_j\hat{p}_k=\frac{i\hbar}{2}\epsilon_{ijk}(\hat{a}_j\hat{a}^\dag_k
+\hat{a}^\dag_k\hat{a}_j)\\
    ~
    &=-i\hbar\epsilon_{ijk}\hat{a}^\dag_j\hat{a}_k-\frac{\hbar}{2}\mu\omega\epsilon_{ijk}\Theta_{jk}.
\end{array}
\end{equation}
The first term is identical in form to the commutative case.
However, it includes non-commutative effects in the expressions of
$\hat{a}_k$ and $\hat{a}^\dag_j$:
$$-i\hbar\epsilon_{ijk}\hat{a}^\dag_j\hat{a}_k=-i\hbar\epsilon_{ijk}
\{\alpha^2 a^\dag_j a_k +\frac{i}{2}\mu\omega (\Theta_{kl}a^\dag_j
a_l -\Theta_{jl}a^\dag_l a_k)
+\frac{\mu^2\omega^2}{4\alpha^2}\Theta_{jl}\Theta_{km}a^\dag_l a_k
\}.$$ The more important thing is the last term in equation
(\ref{L}). This is a new effect proper to noncommutative space. It
tells us that in noncommutative space, the angular momentum has a
non-zero ``zero-point'' angular momentum.  For a given
noncommutative space ($\Theta$ fixed), this ``zero-point'' angular
momentum depends on the mass and angular frequency, so it can have
a fractional value.

In fact, we should point out that the noncommutative effects
always depend on the coordinates commutator matrix $\Theta$ ,and
the matrix elements depend on the scaling constant $\alpha$. From
(\ref{TB}), (\ref{Eq:cc}) and (\ref{alpha}) we have
\begin{equation}
\label{cc2} \Theta_{il}\Theta_{lj}= -\theta^2\delta_{ij},
\end{equation}
where
\begin{equation}
\label{theta1} \theta =
\frac{2\alpha}{\mu\omega}\sqrt{(1-\alpha^2)},
\end{equation}
 and $\alpha$ should be less or equal to  $1$.

 When $n=2$, we obtain,
\begin{equation}
\label{theta2} \Theta =\left( \begin{array}{cc} 0& \theta\\
                                       -\theta & 0
                                       \end{array}\right)
\end{equation}
Our results here for $2$ dimensional space coincide with the
results in reference\cite{zhang}, if we make the following
correspondence,
\begin{equation}
\label{cr} \alpha\Leftrightarrow
\xi^{-1}=(1+dd'/4)^{-\frac{1}{2}},
\end{equation}
and
\begin{equation}
\label{cr2}
\begin{array}{l}
 \Theta_{ij}\Leftrightarrow\xi^{-2}\Lambda^{-2}_{NC}d \epsilon_{ij},\\
 \bar{\Theta}_{ij}\Leftrightarrow\xi^{-2}\Lambda^{2}_{NC} d' \epsilon_{ij}.
\end{array}
\end{equation}
Where $\Lambda_{NC}$ is the noncommutative energy scale. When
$d'=0$,  led to $\alpha=1$ and also $\bar{\Theta}=0$.

When $n=3$,  equation (\ref{cc2}) has three solutions, which read,
\begin{equation}
\label{p=3,1} i):~~ \Theta =\left(
\begin{array}{ccc}
  0&\theta&0\\
  -\theta&0&0\\
  0&0&0
\end{array}\right),~
ii):~\Theta =\left(
\begin{array}{ccc}
  0&0&\theta\\
  0&0&0\\
 -\theta&0&0
\end{array}\right),~
iii):~~\Theta =\left(
\begin{array}{ccc}
  0&0&0\\
  0&0&\theta\\
0&-\theta&0
\end{array}\right).
\end{equation}
These solutions tell us that in our case the noncommutative 3
dimensional space for harmonic oscillators is reducible and, in
fact, it is a direct sum of a $1$-dimensional space and a
$2$-dimensional noncommutative space where the two subspaces are
commutating.

When $n=4$, the equation (\ref{cc2}) has six solutions which can be
classified into three classes. The first class includes two
solutions which have no free parameters. The two solutions for
matrix $\Theta$ are give by
\begin{equation}
\label{p=4,1}  \left(
\begin{array}{cccc}
0&0&0&\theta\\
0&0&\theta&0\\
0&-\theta&0&0\\
-\theta&0&0&0
\end{array}\right),~
\left(
\begin{array}{cccc}
0&0&0&\theta\\
0&0&-\theta&0\\
0&\theta&0&0\\
-\theta&0&0&0
\end{array}\right).
\end{equation}
These solutions correspond to a reducible space which is the
direct sum of two $2$-dimensional independent noncommutative
spaces. The second class has two solutions which contain one free
parameter and have the form
\begin{equation} \label{p=4,2}\left(
\begin{array}{cccc}
  0&\theta_1&\theta_2&0\\
  -\theta_1&0&0&\theta_2\\
-\theta_2&0&0& \theta_1\\
0&-\theta_2&-\theta_1&0
\end{array}\right),~~\left(
\begin{array}{cccc}
  0&\theta_1&\theta_2&0\\
  -\theta_1&0&0&\theta_2\\
-\theta_2&0&0&-\theta_1\\
0&-\theta_2&\theta_1&0
\end{array}\right),
\end{equation}
with $\theta_1$ and $\theta_2$ satisfy $$
\theta_1^2+\theta_2^2=\theta^2 .$$ The solutions of third class
have two free parameters
\begin{equation} \label{p=4,3}\left(
\begin{array}{cccc}
   0&\vartheta_1&\vartheta_2&\vartheta_3\\
   -\vartheta_1&0&\vartheta_3& -\vartheta_2\\
   -\vartheta_2&-\vartheta_3&0&\vartheta_1\\
   -\vartheta_3&\vartheta_2&-\vartheta_1&0
\end{array}\right),~~\left(
\begin{array}{cccc}
   0&\vartheta_1 &\vartheta_2 &\vartheta_3\\
   -\vartheta_1&0&-\vartheta_3& \vartheta_2\\
   -\vartheta_2&\vartheta_3&0&-\vartheta_1\\
   -\vartheta_3&-\vartheta_2&\vartheta_1&0
\end{array}\right),~~
\end{equation}
with $$\vartheta_1^2+\vartheta_2^2+\vartheta_3^2=\theta^2.$$  The
noncommutative spaces for these two class solutions are
irreducible (at least they are non completely reducible for
generic values of the free parameters).

In the last part of this paper, we would like to use the
representation (\ref{Eq:Rep.4}) to find a possible
Schr$\ddot{o}$dinger equation in commutative space which is
equivalent to the the Schr$\ddot{o}$dinger equation in
noncommutative space. To begin with, let us discuss the
Hamiltonian in noncommutative space which should have the
following form
\begin{equation}
\label{H3} \hat{H}=\frac{1}{2\mu}\hat{p}_i\hat{p}_i
+V(\hat{x}_1,...\hat{x}_n)
\end{equation}
Up to the first order in $\Theta$ and $ \bar{\Theta}$, we can
write the noncommutative hamiltonian in terms of commutative
variables as,
\begin{equation}\label{H*}
\begin{array}{ll}
 \hat{H}=&\frac{\alpha^2}{2\mu}p_i p_i +V(\alpha
x_1,\alpha x_2...\alpha x_n) \\
~&-\frac{1}{2\mu}\bar{\Theta}_{ij}x_ip_j-\frac{1}{2\alpha}\frac{\partial
V(x_1,x_2...x_n)}{\partial x_i}\Theta_{ij}p_j .
\end{array}
\end{equation}
In three dimensional space, the Hamiltonian [\ref{H*}] gives the
Schr$\ddot{o}$dinger equation as,
\begin{equation}\label{SchodingerE}
i\hbar\frac{\partial \Psi (\mathbf{x},t)}{\partial t}=
\{-\frac{\alpha^2 \hbar^2}{2\mu}\nabla^2 +V(\alpha\mathbf{x})
-\frac{i\hbar}{2\mu}\mathbf{\tilde{x}}\cdot \nabla
+\frac{i\hbar}{2\alpha}\nabla V \cdot \tilde{\nabla}\}\Psi (
\mathbf{x},t)
\end{equation}
where $\mathbf{\tilde{x}}_i=\bar{\Theta}_{ij}x_j~~
\tilde{\nabla}=\Theta_{ij}\partial_j$. When $\alpha=1$, which
means $\bar{\Theta}=0$, then the transformation relations from
noncommutative space to commutative space become,
\begin{equation}\label{Rep.6}
\begin{array}{l}
\hat{p}_i=p_i\\
\hat{x}_i=x_i-\frac{1}{2}\Theta_{ij}p_j
\end{array}
\end{equation}
Thus, our result reduces to the situation discussed in many papers
\citer{CST,MM}, where the Schr$\ddot{o}$dinger equation  reads,
\begin{equation}
i\hbar\frac{\partial\Psi (\mathbf{x},t)}{\partial
t}=[\frac{p^2}{2\mu}+V(\mathbf{x}-\frac{1}{2}\mathbf{\tilde{p}})]\Psi
(\mathbf{x},t),
\end{equation}
with $\mathbf{\tilde{p}}_i=\Theta_{ij}p_j$.

In summary, we have achieved the following results in this paper.
First, we got a general representation of any dimensional
noncommutative phase space in term of the usual phase space in
quantum mechanics. Then, as a special case, the noncommutative
harmonic oscillator in any dimensions is studied in details, and
we got an constraint for the matrix $\Theta$ related to the
non-commutativity of the space. The results of a few low
dimensional cases are given. At last, we discussed the form of
Schr$\ddot{o}$dinger equation which contains the noncommutative
effects of the phase space. It should be pointed out that most of
our results are not Lorentz invariant, because the time and the
noncommutative spacial coordinates are assumed to commute in this
paper. Nonetheless, the general relation, say equations
(\ref{Eq:Rep.4}) in the first part of this paper, can be used in
the case of noncommutative space-time, where we can set
$\hat{x}_1=ic\hat{t}$.

\paragraph{Acknowledgements.}
This paper was written during Kang Li's visit to High Energy
section of Abdus Salam ICTP, Trieste, Italy and to the Physics
Department of La Plata National University , La Plata, Argentina,
He would like to thank Prof. S.Randjbar-Daemi and Prof. Carlos
Naon for scientific discussions.  This work was supported
partially by the National Nature Science Foundation of China
(90303003) and the Nature Science Foundation of Zhejiang Provence,
China (M103042;102011;102028). The author also recognizes the
support of the Consejo Nacional de Investigaciones
Cient\'{\i}ficas y T\'ecnicas (CONICET), Argentina and the support
of the Third World Academy of Sciences (TWAS).

\end{document}